\documentclass[aps,prc,reprint,floatfix,superscriptaddress,nofootinbib]{revtex4-1}
\usepackage[utf8]{inputenc}
\usepackage{bm}
\usepackage{graphicx}
\usepackage[colorlinks=true, linkcolor=blue, citecolor=red, urlcolor=black, breaklinks=true]{hyperref}

\begin{document}

\title{Probing vorticity structure in heavy-ion collisions by local $\Lambda$
polarization}

\author{Xiao-Liang Xia}
\author{Hui Li}
\author{Zebo Tang}
\author{Qun Wang}

\affiliation{Department of Modern Physics, University of Science and Technology
of China, Hefei, Anhui 230026, China}

\begin{abstract}
We study the local structure of the vorticity field and the $\Lambda$
polarization in Au+Au collisions in the energy range $\sqrt{s_{\mathrm{NN}}}=7.7$--$200$
GeV and Pb+Pb collisions at $\sqrt{s_{\mathrm{NN}}}=2760$ GeV using
A Multi-Phase Transport (AMPT) model. We focus on the vorticity field
arising from the non-uniform expansion of the fireball, which gives
the circular structure of the transverse vorticity $\bm{\omega}_{\perp}=(\omega_{x},\omega_{y})$
around the $z$ direction as well as the quadrupole pattern of the
longitudinal vorticity $\omega_{z}$ in the transverse plane. As a
consequence, the three components of the polarization vector $\mathbf{P}=(P_{x},P_{y},P_{z})$
for $\Lambda$ hyperons show harmonic behaviors as $\mathrm{sgn}(Y)\sin\phi_{p}$,
$-\mathrm{sgn}(Y)\cos\phi_{p}$, and $-\sin(2\phi_{p})$, where $\phi_{p}$
and $Y$ are the azimuthal angle and rapidity in momentum space. These
patterns of the local $\Lambda$ polarization are expected to be tested
in future experiments.
\end{abstract}
\maketitle

\section{Introduction}

In non-central heavy-ion collisions, huge orbital angular momenta
and vorticity fields are produced in strongly coupled quark gluon
plasma (sQGP). They can lead to the hadron polarization and spin alignment
through spin-orbit couplings \cite{Liang:2004ph,Liang:2004xn,Gao:2007bc,Chen:2008wh,Huang:2011ru}
or spin-vorticity couplings \cite{Becattini:2007nd,Becattini:2013fla,Becattini:2016gvu,Fang:2016vpj},
see Refs.~\cite{Wang:2017jpl,Voloshin:2017kqp,Becattini:2017vsh}
for recent reviews. The vorticity-related effects also include some
chiral transport phenomena such as the chiral vortical effect \cite{Son:2009tf}
and the chiral vortical wave \cite{Jiang:2015cva} as well as a change
of the QCD phase diagram \cite{Chen:2015hfc,Chernodub:2016kxh,Huang:2017pqe,Jiang:2016wvv}.

\begin{figure}
\centering\includegraphics[width=0.8\columnwidth]{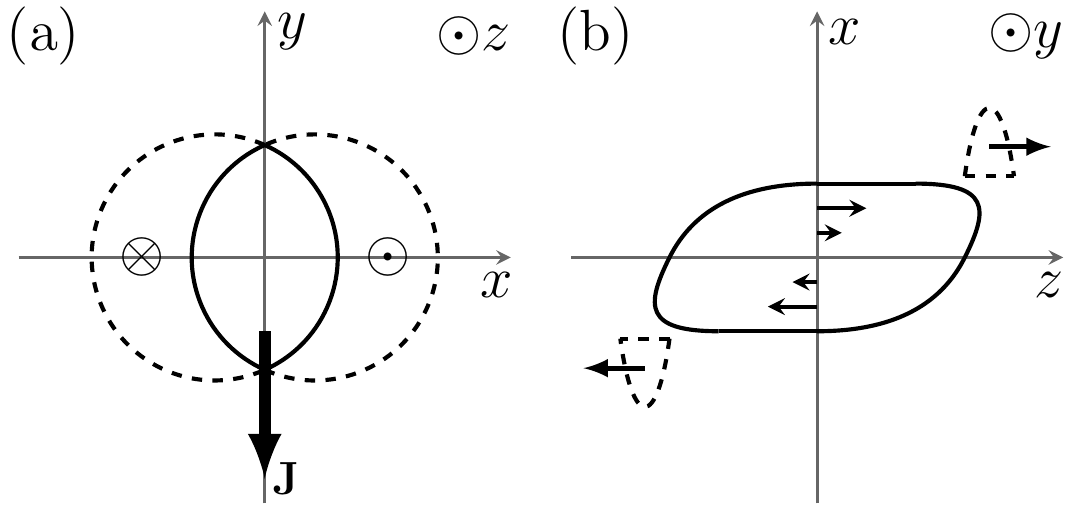}

\caption{\label{fig:collision illustration}Illustration of non-central heavy-ion
collisions in (a) the transverse plane and (b) the reaction plane.
Two nuclei at $(x=\pm b/2,\ y=0)$ move along $\pm z$ direction,
respectively. The global OAM and the net vorticity is along $-y$
direction.}
\end{figure}

The study of the global polarization was initially motivated by the
fact that a huge orbital angular momentum (OAM) is produced in non-central
heavy-ion collisions as shown in Fig.~\ref{fig:collision illustration}(a).
Although such an OAM does not make the sQGP rotating as a rigid body,
it can manifest itself as an initial longitudinal shear flow $\partial_{x}v_{z}>0$
in the fireball as shown in Fig.~\ref{fig:collision illustration}(b).
Then a vorticity field is generated and points to the direction of
the global OAM ($-y$ direction) in average and leads to the global
polarization of hadrons along the same direction.

Recently, the global polarization of $\Lambda$ hyperons in relativistic
heavy-ion collisions has been measured by the STAR Collaboration \cite{STAR:2017ckg}
through their weak decays. The average vorticity of the sQGP has been
extracted to be of order $\omega\sim10^{21}\ \mathrm{s}^{-1}$, the
highest that has ever been found in nature. One feature of the global
polarization is that it decreases with collision energies in the range
of $7.7$--$200$ GeV \cite{STAR:2017ckg}. Several different models
have been used to calculate the vorticity-induced global polarization
of $\Lambda$ hyperons, including hydrodynamics \cite{Karpenko:2016jyx,Xie:2017upb},
A Multi-Phase Transport (AMPT) model with an assumption of local thermodynamical
equilibrium \cite{Li:2017slc,Shi:2017wpk}, AMPT model with the chiral
kinetic equation \cite{Sun:2017xhx}, and the Quark-Gluon-String
Model (QGSM) with anomalous mechanism \cite{Baznat:2017jfj}. The
results of these models show the same energy dependence for the global
$\Lambda$ polarization, which agree with experimental data. For other
studies on vorticity fields or polarizations, see Refs.~\cite{Baznat:2013zx,Baznat:2015eca,Csernai:2013bqa,Csernai:2014ywa,Becattini:2013vja,Becattini:2015ska,Teryaev:2015gxa,Jiang:2016woz,Ivanov:2017dff,Ivanov:2018eej,Deng:2016gyh,Pang:2016igs,Xie:2016fjj,Kolomeitsev:2018svb}.

In a previous paper by some of us \cite{Li:2017slc}, we pointed
out that the global polarization is related to the fireball's tilted
shape in the reaction plane. Due to the faster longitudinal expansion
at higher energies, the fireball or the matter distribution shows
a less tilted shape in mid-rapidity, and thus the net vorticity and
the global polarization are almost vanishing. Such an energy dependence
of the tilted shape can also be seen by the rapidity slope of the
directed flow $dv_{1}/d\eta$ \cite{Voloshin:2017kqp}.

The global polarization is an average effect over the whole volume
of the fireball within the detector's acceptance, so it reflects the
global or net vorticity, which is along the global OAM. However, the
local vorticity field has much richer information than the global
one. In the numerical simulations \cite{Becattini:2015ska,Teryaev:2015gxa,Jiang:2016woz,Li:2017slc,Shi:2017wpk,Ivanov:2017dff,Ivanov:2018eej},
it is observed that $\omega_{y}$ shows a quadrupole pattern in the
reaction plane ($xz$ plane): $\omega_{y}$ is negative and positive
in the regions $xz>0$ and $xz<0$, respectively. This novel structure
is mainly due to the fact that the transverse velocity $|v_{x}|$
decreases with rapidity or $\partial|v_{x}|/\partial|z|<0$ \cite{Jiang:2016woz}.
In addition to the pattern of $\omega_{y}$, a similar quadrupole
structure of $\omega_{z}$ also exists in the transverse plane \cite{Becattini:2017gcx,Karpenko:2017dui,Voloshin:2017kqp}
since the transverse velocity $\mathbf{v}_{\perp}=(v_{x},v_{y})$
is not exactly along the radial direction $\mathbf{e}_{r}$ due to
the fireball's elliptic shape in the transverse plane.

In this paper, we give a systematic analysis of the patterns of the
fluid velocity and vorticity in the fireball produced in heavy-ion
collisions. We find that all components, $\omega_{x}$, $\omega_{y}$,
and $\omega_{z}$, have quadrupole patterns in the $yz$, $xz$, and
$xy$ plane, respectively. These quadrupole patterns all arise from
the fireball expansion not related to the OAM. In order to probe the
quadrupole pattern of the vorticity field, one can separate the whole
momentum space into different regions and measure the average $\Lambda$
polarization in each region separately. Through the numerical simulation
with the AMPT model, we find that the quadrupole patterns of vorticity
fields can lead to a sizable local $\Lambda$ polarization and are
expected to be measured in future experiments.

This paper is organized as follows. We first give a brief introduction
to the polarization induced by vorticity in Sec.~\ref{sec:Basic-definition}.
Then we discuss in Sec.~\ref{sec:Velocity-and-vorticity} how the
fluid velocity and vorticity are distributed in the fireball. The
local vorticity structure can be probed by the local $\Lambda$ polarization.
In Sec.~\ref{sec:Result-and-discussion}, we present the numerical
results for the local $\Lambda$ polarization by the AMPT model. The
purpose of our numerical calculation is to give some typical features
of the local $\Lambda$ polarization, which can be used to probe the
local vorticity distribution in future experiments. A summary of results
is given in the final section.

\section{\label{sec:Basic-definition}Basics on vorticity, spin, and polarization}

In this section, we give a brief introduction to the polarization
induced by vorticity in a hydrodynamical system. In a non-relativistic
fluid, the vorticity can be used to characterize the local rotation
of the fluid,
\begin{equation}
\bm{\omega}=\frac{1}{2}\nabla\times\mathbf{v},\label{eq:vorticity_nr}
\end{equation}
where $\mathbf{v}$ is the fluid velocity as a function of space-time.
Particles with spin degrees of freedom in the vortical fluid are expected
to be polarized in alignment with the vorticity. In local thermodynamical
equilibrium, the ensemble average of the spin vector for the spin-1/2
particle is given by $\mathbf{S}=\mathrm{tr}(\rho\widehat{\mathbf{S}})$,
where $\widehat{\mathbf{S}}=\bm{\sigma}/2$ is the spin operator with
$\bm{\sigma}$ being Pauli matrices and $\rho\sim\exp(\widehat{\mathbf{S}}\cdot\bm{\omega}/T)$
is the spin density matrix with the temperature $T$. One can obtain
$\mathbf{S}$ as
\begin{equation}
\mathbf{S}=\frac{1}{2}\tanh\left(\frac{\omega}{T}\right)\hat{\bm{\omega}}\simeq\frac{\bm{\omega}}{4T},\label{eq:spin_vector_nr}
\end{equation}
where $\hat{\bm{\omega}}=\bm{\omega}/|\bm{\omega}|$ is the direction
of $\bm{\omega}$. The polarization vector for the spin-1/2 particle
is defined as
\begin{equation}
\mathbf{P}\equiv2\mathbf{S}=\tanh\left(\frac{\omega}{T}\right)\hat{\bm{\omega}}\simeq\frac{\bm{\omega}}{2T},\label{eq:polarization_vector_nr}
\end{equation}
where the factor $2$ is introduced to normalize the polarization
magnitude to unity.

Since the high-energy heavy-ion collision is a relativistic system,
the above equations should be generalized to the relativistic ones.
As argued in Refs.~\cite{Becattini:2007nd,Becattini:2013fla}, the
quantity that is related to the polarization is the thermal vorticity
tensor $\varpi_{\mu\nu}$ defined by
\begin{equation}
\varpi_{\mu\nu}=\frac{1}{2}\left(\partial_{\nu}\beta_{\mu}-\partial_{\mu}\beta_{\nu}\right),\label{eq:vorticity}
\end{equation}
where $\beta^{\mu}=u^{\mu}/T$ with $u^{\mu}=\gamma(1,\mathbf{v})$
being the fluid four-velocity and $\gamma=1/\sqrt{1-\mathbf{v}^{2}}$
the Lorentz factor. The vorticity tensor $\varpi_{\mu\nu}$ can be
decomposed into two groups of components as
\begin{eqnarray}
\bm{\varpi}_{T} & = & \left(\varpi_{0x},\varpi_{0y},\varpi_{0z}\right)=\frac{1}{2}\left[\nabla\left(\frac{\gamma}{T}\right)+\partial_{t}\left(\frac{\gamma\mathbf{v}}{T}\right)\right],\label{eq:vorticity_T}\\
\bm{\varpi}_{S} & = & \left(\varpi_{yz},\varpi_{zx},\varpi_{xy}\right)=\frac{1}{2}\nabla\times\left(\frac{\gamma\mathbf{v}}{T}\right).\label{eq:vorticity_S}
\end{eqnarray}
In the Boltzmann limit and linear order of $\varpi_{\mu\nu}$, the
spin vector is given by \cite{Li:2017slc}
\begin{equation}
\mathbf{S}\left(x,p\right)=\frac{1}{4m}\left(E_{p}\bm{\varpi}_{S}+\mathbf{p}\times\bm{\varpi}_{T}\right).\label{eq:spin_vector}
\end{equation}
where $E_{p}$, $\mathbf{p}$, and $m$ are the energy, momentum,
and mass of the particle. In the numerical simulations, $\mathbf{S}$
in Eq.~(\ref{eq:spin_vector}) is usually calculated in the center
of mass frame of A+A collisions, while in experiments the polarization
is measured in the $\Lambda$'s rest frame by the angular distribution
of the proton in $\Lambda$'s weak decay. To obtain the spin vector
$\mathbf{S}^{*}$ in the $\Lambda$'s rest frame from $\mathbf{S}$
in the calculational frame, one uses the Lorentz transformation
\begin{equation}
\mathbf{S}^{*}=\mathbf{S}-\frac{\mathbf{p}\cdot\mathbf{S}}{E_{p}\left(m+E_{p}\right)}\mathbf{p}.\label{eq:Lorentz_boost}
\end{equation}
In this case, the $\Lambda$ polarization vector is given by
\begin{equation}
\mathbf{P}=2\mathbf{S}^{*},\label{eq:polarization_vector}
\end{equation}
corresponding to the non-relativistic one in Eq.~(\ref{eq:polarization_vector_nr}).

Both Eqs.~(\ref{eq:spin_vector_nr}), (\ref{eq:polarization_vector_nr})
and their relativistic generalizations, Eqs.~(\ref{eq:spin_vector})--(\ref{eq:polarization_vector}),
relate the polarization to the vorticity field where and when a particle
such as $\Lambda$ is formed at one space-time point. Therefore the
local structure of the vorticity can be probed by measuring the local
$\Lambda$ polarization. This is the general idea we will follow throughout
this study.

\section{\label{sec:Velocity-and-vorticity}Fluid velocity and vorticity fields
and polarization distributions}

We now discuss the fluid velocity and vorticity fields. In order to
give an intuitive picture, our discussion is based on Eqs.~(\ref{eq:vorticity_nr})--(\ref{eq:polarization_vector_nr})
using non-relativistic fields $\mathbf{v}$ and $\bm{\omega}$, while
the numerical calculations for polarizations in the next section are
based on relativistic quantities in Eqs.~(\ref{eq:vorticity})--(\ref{eq:polarization_vector})
instead. The non-relativistic polarization defined by Eqs.~(\ref{eq:vorticity_nr})--(\ref{eq:polarization_vector_nr})
should have the same feature as the relativistic one by Eqs.~(\ref{eq:vorticity})--(\ref{eq:polarization_vector}).

Throughout this paper, we use the coordinate system shown in Fig.~\ref{fig:collision illustration},
where two nuclei at $(x=\pm b/2,\ y=0)$ in the transverse plane move
along the $\pm z$ direction, respectively, and the global OAM $\mathbf{J}$
is along the $-y$ direction. Due to the OAM of the fireball, a net
vorticity field is expected to form whose direction is pointing to
the OAM in average. In the following we formally denote such a net
vorticity as $\langle\omega_{y}\rangle$, where $\langle\rangle$
means the average over space weighted by the matter density. This
net vorticity leads to a global polarization $\mathbf{P}^{G}$ along
the direction of the OAM.

\begin{figure}[t]
\centering\includegraphics[width=1\columnwidth]{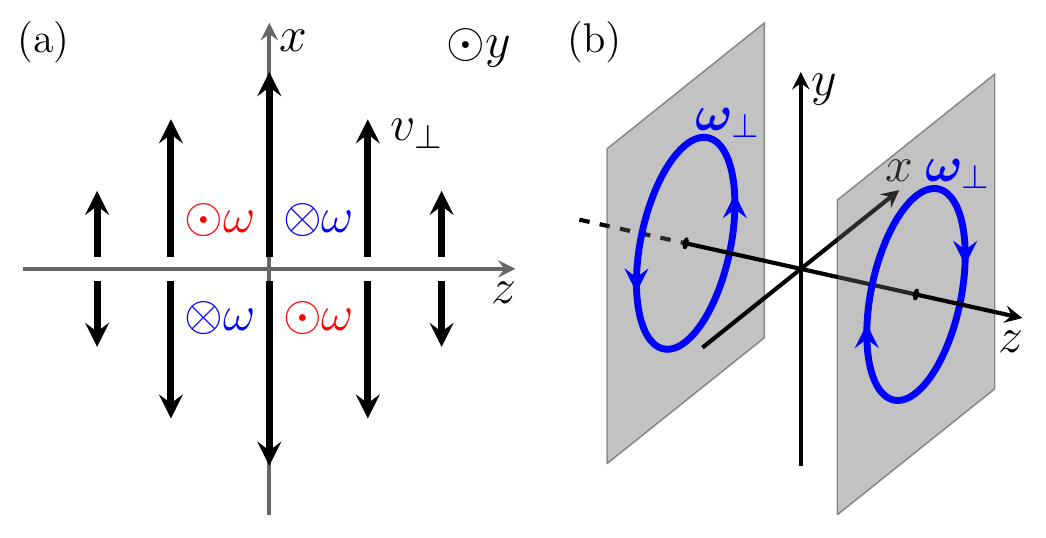}

\caption{\label{fig:structure_2}(a) Schematic illustration of the quadrupole
pattern of $\omega_{y}$ generated from $\partial_{z}v_{\perp}$ in
the reaction plane, where the vorticity is along the $-y$ direction
($\otimes$) in the $xz>0$ quadrants and the $y$ direction ($\odot$)
in the $xz<0$ quadrants. (b) A three-dimensional view of the circular
structure of the transverse vorticity $\bm{\omega}_{\perp}=(\omega_{x},\omega_{y})$.}
\end{figure}

Apart from the net vorticity originated from the OAM, vorticity can
also be generated from the fireball's non-uniform expansion as illustrated
by Fig.~\ref{fig:structure_2}, whose pattern is very different from
the net vorticity (along $-y$ direction). For simplicity, let us
consider an isotropic transverse velocity field $\mathbf{v}_{\perp}=(v_{x},v_{y})$
in the following form
\begin{equation}
\mathbf{v}_{\perp}=v_{\perp}(r,z)\mathbf{e}_{r},\label{eq:transverse velocity}
\end{equation}
where $r$ and $z$ are the transverse radius and the longitudinal
coordinate, and $\mathbf{e}_{r}$ is the unit vector along the radial
direction in the transverse plane. Then from Eq.~(\ref{eq:vorticity_nr}),
the transverse vorticity filed $\bm{\omega}_{\perp}=(\omega_{x},\omega_{y})$
is given by
\begin{equation}
\bm{\omega}_{\perp}=\frac{1}{2}\partial_{z}v_{\perp}(r,z)\mathbf{e}_{\phi},\label{eq:circular vorticity}
\end{equation}
where $\mathbf{e}_{\phi}=(-\sin\phi,\cos\phi,0)$ is the unit vector
along the azimuthal direction with $\phi$ being the azimuthal angle
with respect to the $x$ axis. If the fluid is the Bjorken-type with
the longitudinal boost invariance that $v_{\perp}$ is independent
of $z$, then $\bm{\omega}_{\perp}$ is zero. However, in realistic
collisions the longitudinal boost invariance is violated since the
matter is not uniformly distributed in space, which can give rise
to a nonzero vorticity. Note that the energy or matter is mostly deposited
at $z=0$, the pressure-driven transverse velocity $v_{\perp}$ should
be the largest at $z=0$ and decrease with $|z|$ as shown in Fig.~\ref{fig:structure_2}(a).
Then with the gradients $\partial v_{\perp}/\partial|z|<0$, one can
see that $\bm{\omega}_{\perp}$ in Eq.~(\ref{eq:circular vorticity})
has a circular structure: $\bm{\omega}_{\perp}$ is along $-\mathbf{e}_{\phi}$
(clockwise) and $\mathbf{e}_{\phi}$ (counter-clockwise) in the $z>0$
and $z<0$ regions respectively as shown in Fig.~\ref{fig:structure_2}(b).
In terms of the components $\omega_{x}$ and $\omega_{y}$, they have
the quadrupole structures: $\omega_{x}>0$ ($\omega_{x}<0$) in the
$yz>0$ ($yz<0$) quadrants and $\omega_{y}>0$ ($\omega_{y}<0$)
in the $xz<0$ ($xz>0$) quadrants.

\begin{figure}[t]
\centering\includegraphics[width=1\columnwidth]{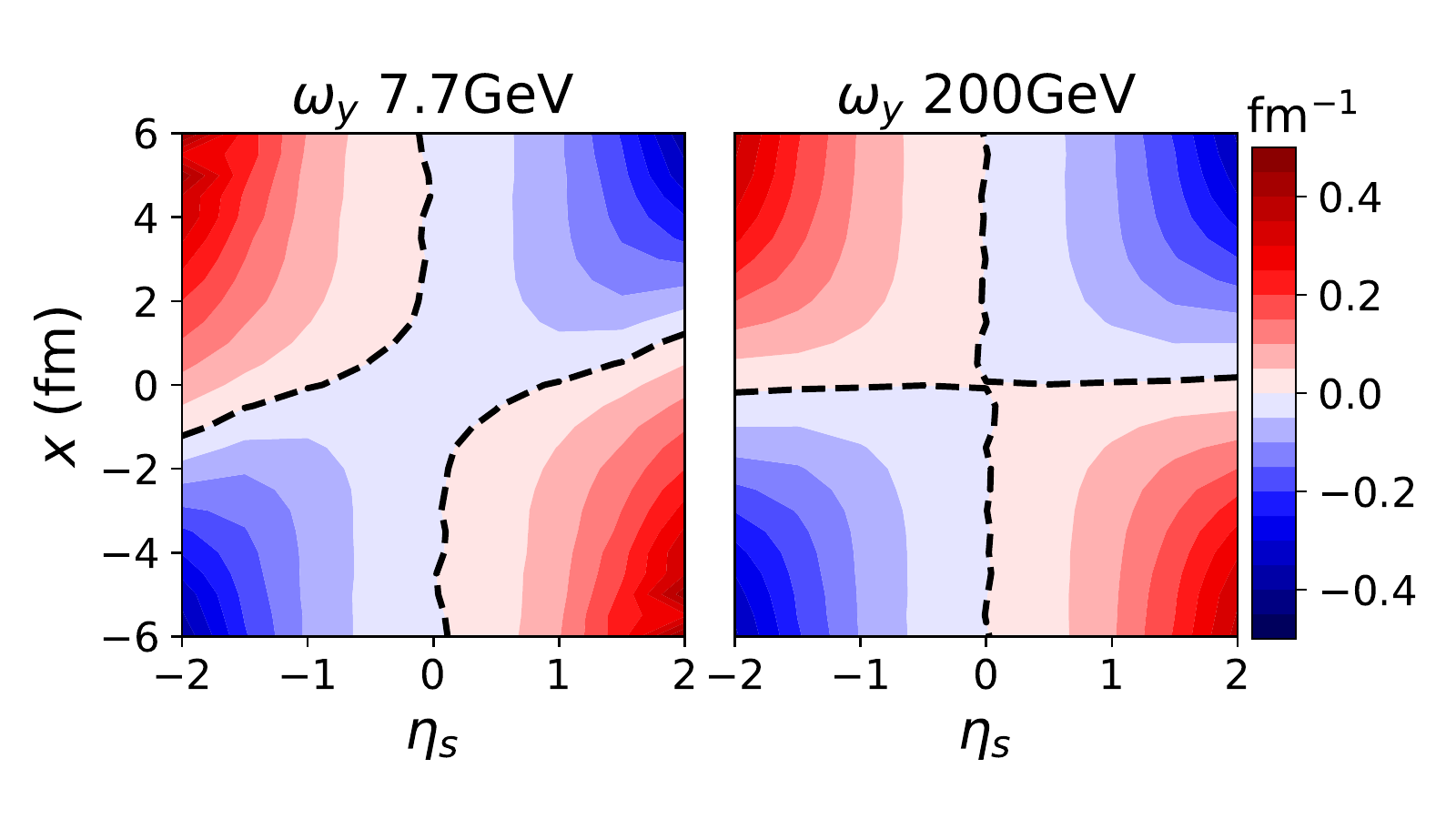}

\caption{\label{fig:omega_y}The vorticity component $\omega_{y}$ in the reaction
plane ($x\eta_{s}$ plane at $y=0$) at time $t=5\ \mathrm{fm}/c$
in $20$--$30\%$ central Au+Au collisions at $\sqrt{s_{\mathrm{NN}}}=7.7$
(left) and $200$ GeV (right). The black dashed lines represent the
contour where $\omega_{y}=0$.}
\end{figure}

\begin{figure}[t]
\centering\includegraphics[width=1\columnwidth]{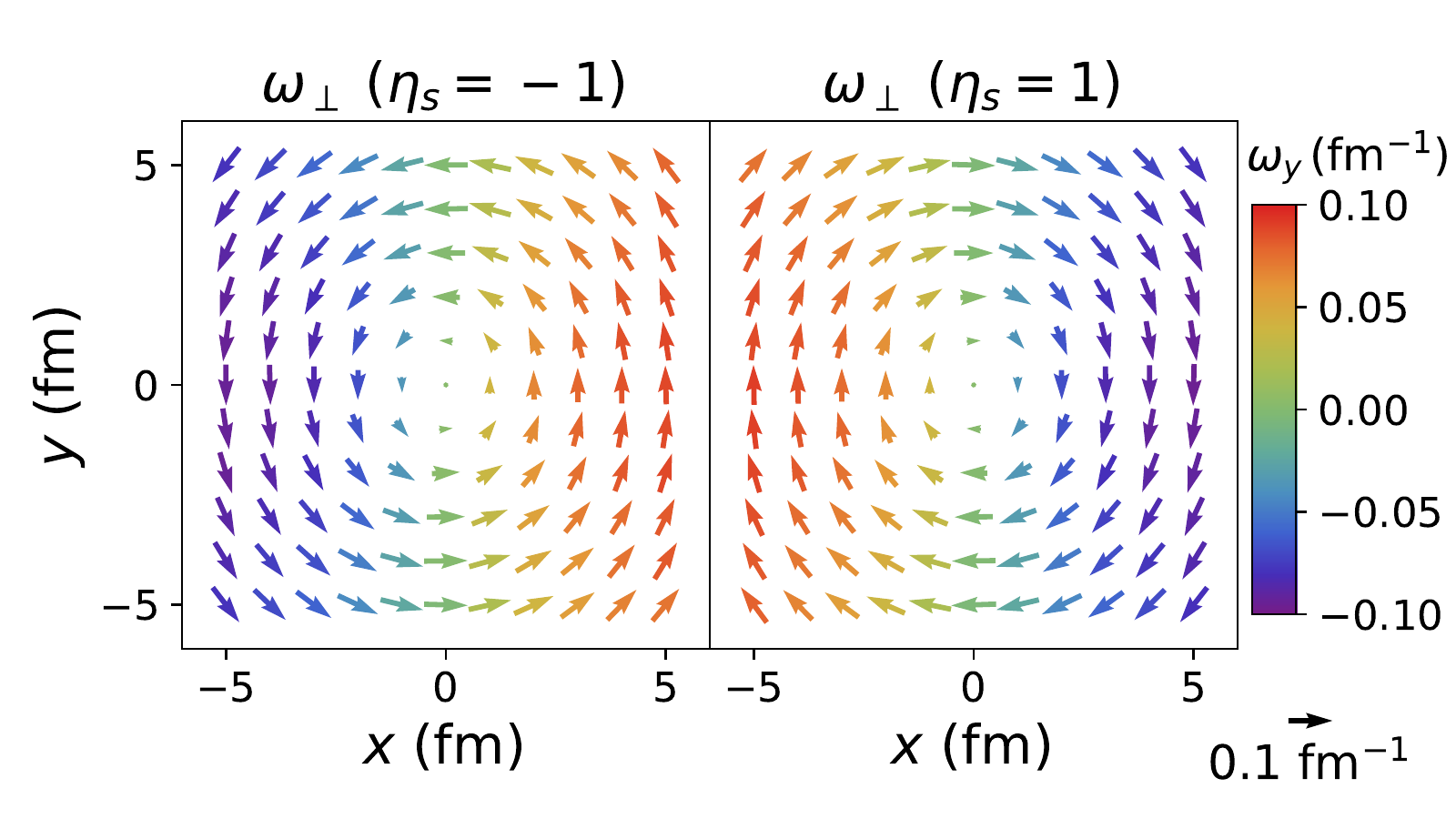}

\caption{\label{fig:omega_perp}The distribution of the transverse vorticity
$\bm{\omega}_{\perp}=(\omega_{x},\omega_{y})$ in the transverse plane
at longitudinal positions $\eta_{s}=-1$ (left) and $\eta_{s}=1$
(right) at time $t=5\ \mathrm{fm}/c$ in $20$--$30\%$ central Au+Au
collisions at $\sqrt{s_{\mathrm{NN}}}=200$ GeV. The color represents
the value of the component $\omega_{y}$.}
\end{figure}

\begin{table*}[t]
\centering

\caption{\label{table:vorticity_pattern}The sources and $\Lambda$ polarization
effects of different vorticity patterns.}

\begin{ruledtabular}
\begin{tabular}{lll}
Vorticity pattern  & Source  & Effect on $\Lambda$ polarization\tabularnewline
\hline
net vorticity $\langle\omega_{y}\rangle<0$  & global OAM or fireball's tilted shape  & global polarization $P_{y}^{G}$\tabularnewline
circular structure of $\bm{\omega}_{\perp}$  & longitudinal dependence of transverse velocity  & local polarization, see $P_{x}$ and $P_{y}$ in Eq.~(\ref{eq:Fourier})\tabularnewline
quadrupole structure of $\omega_{z}$  & anisotropic transverse velocity (elliptic flow)  & local polarization, see $P_{z}$ in Eq.~(\ref{eq:Fourier})\tabularnewline
\end{tabular}
\end{ruledtabular}

\end{table*}

We show the quadrupole or circular pattern of the transverse vorticity
obtained in numerical calculations with the AMPT model in Figs.~\ref{fig:omega_y}
and \ref{fig:omega_perp}, which correspond to the illustrations in
Figs.~\ref{fig:structure_2}(a) and \ref{fig:structure_2}(b), respectively.

Figure \ref{fig:omega_y} shows the $y$ component of the vorticity
in the $x\eta_{s}$ plane at $y=0$, where $\eta_{s}=(1/2)\log[(t+z)/(t-z)]$
is the space-time rapidity. Here the vorticity field is shown at the
time $t=5\ \mathrm{fm}/c$ in $20$--$30\%$ central Au+Au collisions
at $\sqrt{s_{\mathrm{NN}}}=7.7$ and $200$ GeV. We see that $\omega_{y}$
at $200$ GeV has a nearly prefect quadrupole structure: $\omega_{y}$
is an odd function of both $x$ and $\eta_{s}$. This pattern is consistent
with what we expect in Fig.~\ref{fig:structure_2}(a). As for $7.7$
GeV, $\omega_{y}$ is not an odd function. In particular, we see that
$\omega_{y}<0$ in the central region $x\simeq\eta_{s}\simeq0$. Such
a deviation from the odd function comes from the fireball's tilted
geometry in the reaction plane in non-central collisions as shown
in Fig.~\ref{fig:collision illustration}(b). We can regard the pattern
of $\omega_{y}$ as a sum of two different vorticity patterns: the
net vorticity generated from the fireball's tilted shape and the quadrupole
one from the fireball's non-uniform expansion. The net vorticity pattern
has an obvious energy dependence. At $200$ GeV, the contribution
from the net vorticity is very small since the fireball is less tilted
in mid-rapidity at higher energy due to its faster longitudinal expansion
\cite{Li:2017slc,Xia:2017iav}. In contrast the quadrupole vorticity
has the same magnitude at $7.7$ and $200$ GeV. See Refs.~\cite{Becattini:2015ska,Teryaev:2015gxa,Jiang:2016woz,Li:2017slc,Shi:2017wpk,Ivanov:2017dff,Ivanov:2018eej}
for other calculations of the quadrupole structure of $\omega_{y}$.

Figure \ref{fig:omega_perp} shows the distribution of $\bm{\omega}_{\perp}=(\omega_{x},\omega_{y})$
as functions of $x$ and $y$ at two values of space-time rapidity
$\eta_{s}=-1$ and $1$. Here the vorticity field is shown at the
time $t=5\ \mathrm{fm}/c$ in $20$--$30\%$ central Au+Au collisions
at $\sqrt{s_{\mathrm{NN}}}=200$ GeV for instance. We see that $\bm{\omega}_{\perp}$
has a circular structure with opposite orientations in the $\eta_{s}>0$
and $\eta_{s}<0$ regions. This pattern is consistent with what we
expect in Fig.~\ref{fig:structure_2}(b). The behavior that the magnitude
of $\bm{\omega}_{\perp}$ increases with the transverse radius $r$
can be understood by the increase of $v_{\bot}(r,z)$ with $r$. The
circular pattern of $\bm{\omega}_{\perp}$ has also been observed
in Ref.~\cite{Baznat:2015eca}.

Besides the transverse component $\bm{\omega}_{\perp}$, the longitudinal
component $\omega_{z}$ also has a non-vanishing local distribution.
Due to the anisotropic flow, $\mathbf{v}_{\perp}$ in non-central
collisions is not along the radial direction $\mathbf{e}_{r}$, this
gives rise to the inequality of $\partial_{x}v_{y}$ and $\partial_{y}v_{x}$,
and then a non-vanishing $\omega_{z}$ with the quadrupole pattern
in the transverse plane: $\omega_{z}$ are along the opposite directions
in the regions $\sin(2\phi)>0$ and $\sin(2\phi)<0$ \cite{Becattini:2017gcx,Voloshin:2017kqp}.

In experiments, one can measure the local $\Lambda$ polarization
to probe the quadrupole or circular pattern of the vorticity field.
Due to the collective expansion of the fireball, the space information
of the vorticity field can be reflected by the local $\Lambda$ polarization
as functions of $\phi_{p}$ and $Y$, where $\phi_{p}$ is the azimuthal
angle of $\Lambda$'s momentum with respect to the reaction plane
and $Y=(1/2)\log[(E_{p}+p_{z})/(E_{p}-p_{z})]$ is the momentum rapidity.
From the circular structure of the transverse vorticity $\bm{\omega}_{\perp}$
in Eq.~(\ref{eq:circular vorticity}) and the quadrupole pattern
of the longitudinal vorticity $\omega_{z}$, to the leading order
of the Fourier decomposition, we expect that the polarization vector
$\mathbf{P}=(P_{x},P_{y},P_{z})$ for the $\Lambda$ hyperon has the
following harmonic behavior:
\begin{eqnarray}
P_{x}(\phi_{p},Y) & = & F_{x}\mathrm{sgn}(Y)\sin\phi_{p},\nonumber \\
P_{y}(\phi_{p},Y) & = & -F_{y}\mathrm{sgn}(Y)\cos\phi_{p},\nonumber \\
P_{z}(\phi_{p},Y) & = & -F_{z}\sin(2\phi_{p}),\label{eq:Fourier}
\end{eqnarray}
where $F_{x}$, $F_{y}$, and $F_{z}$ are the Fourier coefficients,
which are all positive, and $\mathrm{sgn}(Y)$ denotes the sign of
$Y$ coming from the opposite circular orientations of $\bm{\omega}_{\perp}$
at $\eta_{s}>0$ and $\eta_{s}<0$ as shown in Fig.~\ref{fig:structure_2}(b)
and Fig.~\ref{fig:omega_perp}.

In summary of this section, there are three different vorticity patterns
as listed in Table~\ref{table:vorticity_pattern}. They are the net
vorticity $\langle\omega_{y}\rangle$ from the fireball's tilted shape
in the reaction plane, the circular transverse vorticity $\bm{\omega}_{\perp}$
from the longitudinal dependence of the transverse velocity, and the
quadrupole structure of $\omega_{z}$ from the anisotropic transverse
velocity. These three vorticity patterns can lead to the global polarization
$P_{y}^{G}$, the circular polarization $(P_{x},P_{y})$ in the transverse
directions and the local polarization $P_{z}$ in the longitudinal
direction, respectively. Since the three vorticity patterns are from
different sources, their effects on $\Lambda$ polarization could
have different energy and centrality behaviors, which are studied
in the next section.

\section{\label{sec:Result-and-discussion}Numerical results and discussions}

In this section, we show the numerical results of the local $\Lambda$
polarization using the string-melting version of the AMPT model \cite{Lin:2004en}
as the event generator. In this model, the collision participants
are converted to partons, which are allowed to interact by two-body
elastic scatterings. In this partonic phase the collective flow velocity
and vorticity are generated, and we calculate the thermal vorticity
$\varpi_{\mu\nu}$ in Eq.~(\ref{eq:vorticity}) by the same coarse-grain
method as in Ref.~\cite{Li:2017slc}. At the end of the partonic
phase, the $\Lambda$ hyperons are produced by a coalescence mechanism.
Then their polarizations are calculated with Eqs.~(\ref{eq:spin_vector})--(\ref{eq:polarization_vector})
with the values of $\varpi_{\mu\nu}$ at the space-time point at which
the $\Lambda$ hyperons are produced.

We run simulations of Au+Au collisions at energies $\sqrt{s_{\mathrm{NN}}}=7.7$,
$11.5$, $14.5$, $19.6$, $27$, $39$, $62.4$, $200$ GeV and also
Pb+Pb collisions at $2760$ GeV. For each collision energy, $5\times10^{5}$
events are generated with varying impact parameter $b$ limited to
the range $0$--$25$ fm. These events are classified into different
centrality bins $0$--$10\%$, $10$--$20\%$, $\dots$ according
to the charged particle multiplicities. For each collision energy
and centrality, we calculate the polarizations of all the $\Lambda$
hyperons in the rapidity region $|Y|<1$.

To probe the vorticity structure, we group all $\Lambda$ hyperons
into several bins by their azimuthal angle $\phi_{p}$ and the sign
of $Y$. Then we calculate the average of the polarization vector
for $\Lambda$ in each bin. The result $\mathbf{P}(\phi_{p},Y)$ is
obtained as a function of $\phi_{p}$ and $Y$. To test Eq.~(\ref{eq:Fourier}),
we define the following quantities:
\begin{eqnarray}
\left\langle P_{x}\mathrm{sgn}(Y)\right\rangle  & = & \frac{P_{x}(\phi_{p},Y>0)-P_{x}(\phi_{p},Y<0)}{2},\nonumber \\
\left\langle P_{y}\mathrm{sgn}(Y)\right\rangle  & = & \frac{P_{y}(\phi_{p},Y>0)-P_{y}(\phi_{p},Y<0)}{2},\nonumber \\
\left\langle P_{z}\right\rangle  & = & \frac{P_{z}(\phi_{p},Y>0)+P_{z}(\phi_{p},Y<0)}{2}.\label{eq:weighted_average_polarization}
\end{eqnarray}
Here the averages in the first two lines are taken with weight $\mathrm{sgn}(Y)$.
In this way, the global polarization $\mathbf{P}^{G}$ along $-y$
direction in both $Y>0$ and $Y<0$ region is removed, but the circular
polarizations $P_{x}$ and $P_{y}$ in Eq.~(\ref{eq:Fourier}) survive.
Then using Eq.~(\ref{eq:weighted_average_polarization}), we can
focus on the effects from the circular or quadrupole patterns of the
local vorticity. Note that the three quantities in Eq.~(\ref{eq:weighted_average_polarization})
are functions of $\phi_{p}$ only.

\begin{figure}[h]
\centering\includegraphics[width=0.9\columnwidth]{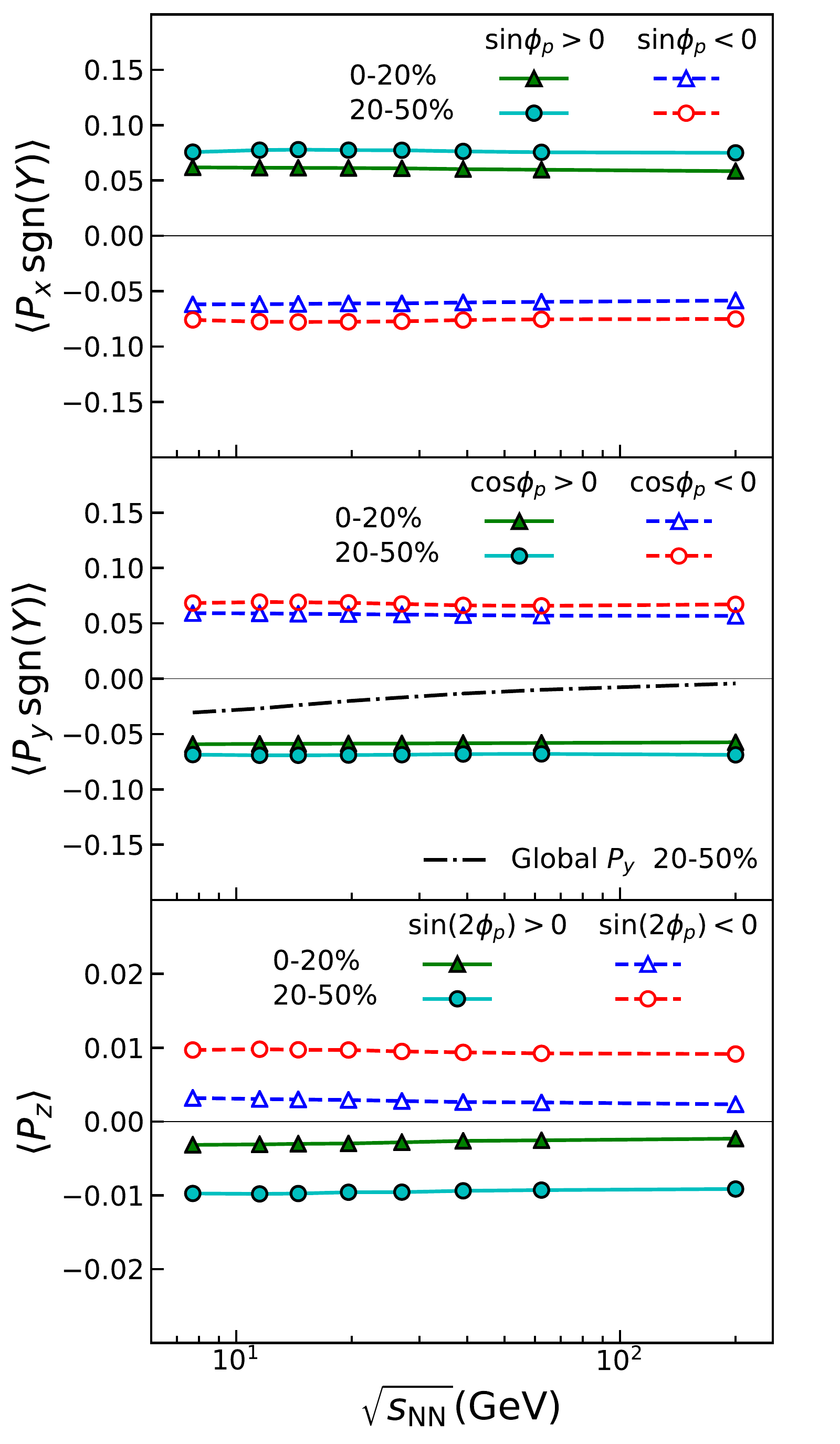}

\caption{\label{fig:P_energy}The average polarizations $\langle P_{x}\mathrm{sgn}(Y)\rangle$,
$\langle P_{y}\mathrm{sgn}(Y)\rangle$, and $\langle P_{z}\rangle$
for the $\Lambda$ hyperons with different signs of $\sin\phi_{p}$,
$\cos\phi_{p}$, and $\sin(2\phi_{p})$ as functions of collision
energies in the range $7.7$--$200$ GeV in $0$--$20\%$ (triangle)
and $20$--$50\%$ (circle) central Au+Au collisions. The black dot-dashed
line shown in the middle panel refers to the global polarization of
$\Lambda$ hyperons in the centrality bin $20$--$50\%$ in mid-rapidity
region $|Y|<1$.}
\end{figure}

\begin{figure*}
\centering\includegraphics[width=0.9\columnwidth]{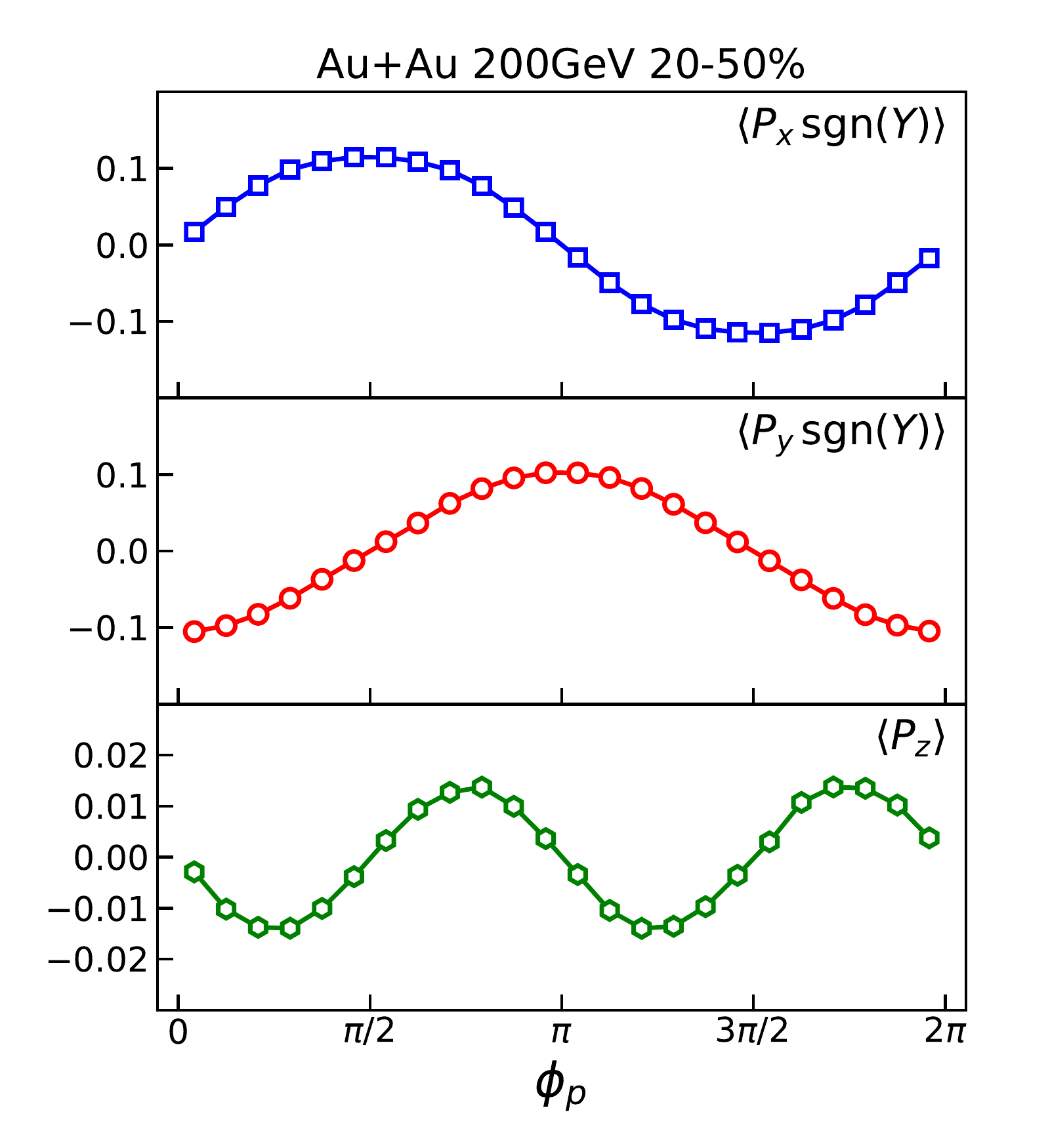}\includegraphics[width=0.9\columnwidth]{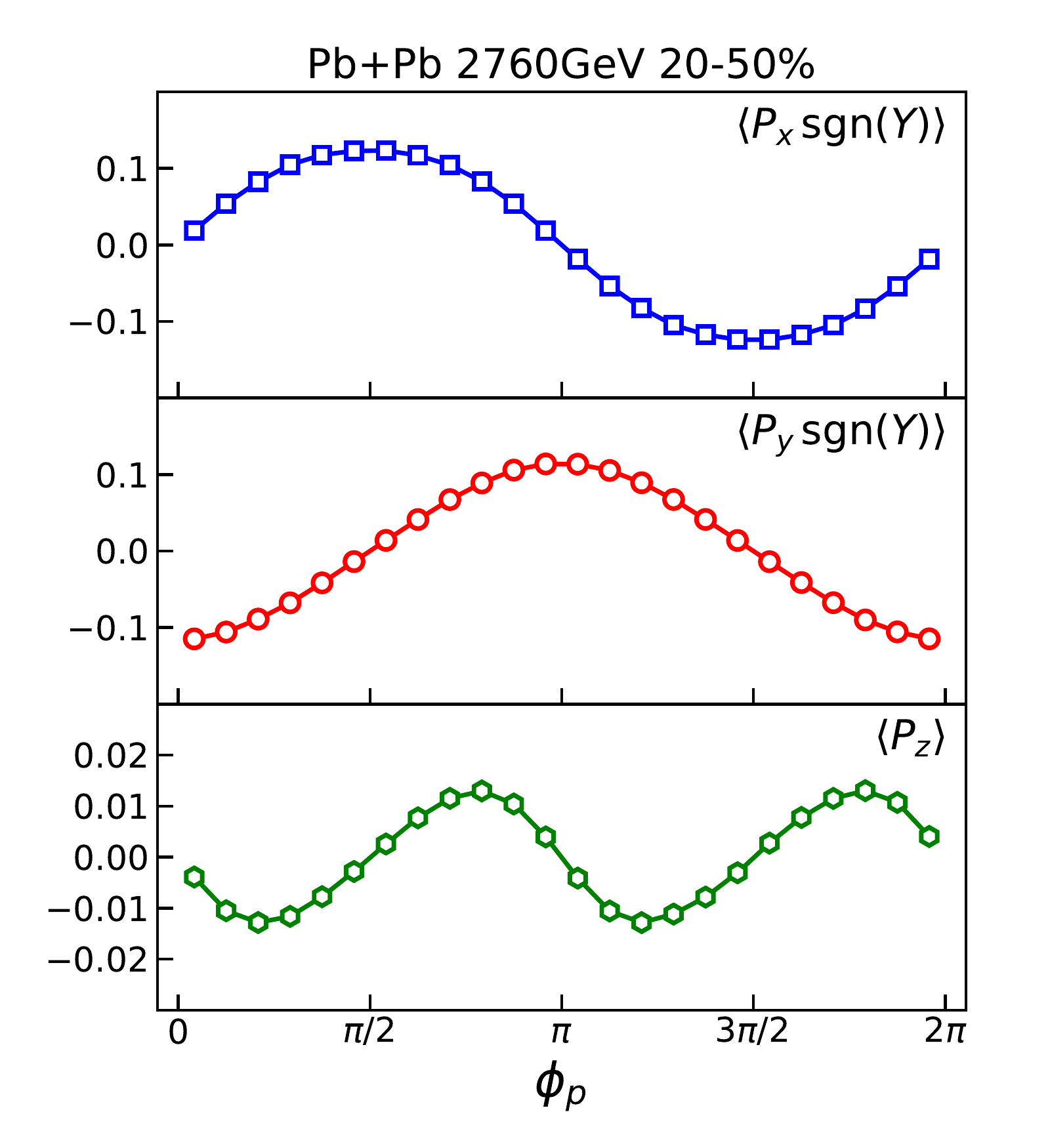}

\caption{\label{fig:P_phi}The average polarizations $\langle P_{x}\mathrm{sgn}(Y)\rangle$,
$\langle P_{y}\mathrm{sgn}(Y)\rangle$, and $\langle P_{z}\rangle$
for $\Lambda$ as functions of azimuthal angle $\phi_{p}$ in $20$--$50\%$
central Au+Au collisions at $200$ GeV (left) and Pb+Pb collisions
at $2760$ GeV (right).}
\end{figure*}

\begin{figure}[t]
\centering\includegraphics[width=0.9\columnwidth]{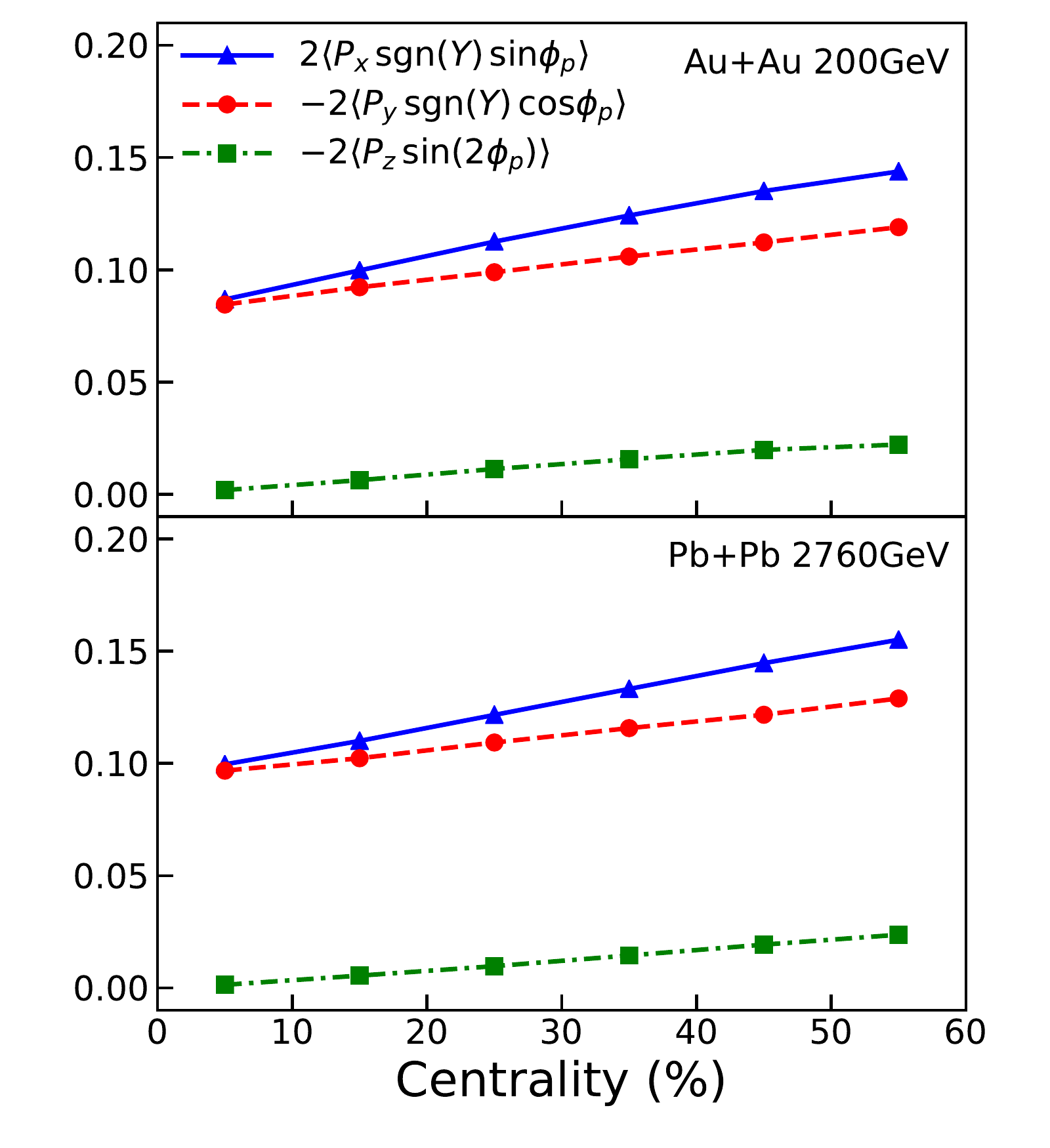}

\caption{\label{fig:P_cen}The Fourier coefficients in Eq.~(\ref{eq:Fourier_coefficient})
as functions of the centrality at $200$ GeV for Au+Au collisions
(top) and $2760$ GeV for Pb+Pb collisions (bottom).}
\end{figure}

Figure \ref{fig:P_energy} shows the results of $\langle P_{x}\mathrm{sgn}(Y)\rangle$,
$\langle P_{y}\mathrm{sgn}(Y)\rangle$, and $\langle P_{z}\rangle$
for the $\Lambda$ hyperons in different regions of $\phi_{p}$ as
functions of collision energies in the range $7.7$--$200$ GeV in
$0$--$20\%$ and $20$--$50\%$ central Au+Au collisions. Here
the azimuthal angle is divided into different regions by the signs
of $\sin\phi_{p}$, $\cos\phi_{p}$, and $\sin(2\phi_{p})$ for the
calculations of $P_{x}$, $P_{y}$, and $P_{z}$, respectively. We
can see that $\langle P_{x}\mathrm{sgn}(Y)\rangle$ is positive (negative)
in region $\sin\phi_{p}>0$ ($<0$) and $\langle P_{y}\mathrm{sgn}(Y)\rangle$
is negative (positive) in region $\cos\phi_{p}>0$ ($<0$), which
are consistent with the circular structure of the transverse vorticity.
Also we see $\langle P_{z}\rangle$ is negative (positive) in region
$\sin(2\phi_{p})>0$ ($<0$), which is also consistent with the quadrupole
pattern of $\omega_{z}$ in the transverse plane.

For comparison, we also show the global polarization $\mathbf{P}^{G}$
of $\Lambda$ by the black dot-dashed line in the middle panel of
Fig.~\ref{fig:P_energy}. The global polarization $\mathbf{P}^{G}$
is the effect from the net vorticity $\langle\omega_{y}\rangle$.
It is calculated by taking an average over all $\Lambda$ hyperons
without dividing them into bins by $\phi_{p}$ and $Y$. From the
reflection symmetry of the fireball, one can prove that only the $P_{y}^{G}$
component (along the OAM) is non-vanishing. As discussed in Sec.~\ref{sec:Velocity-and-vorticity},
the global polarization $P_{y}^{G}$ and the net vorticity $\langle\omega_{y}\rangle$
are originated from the fireball's tilted shape in the reaction plane.
Since the fireball is less tilted in mid-rapidity at higher energies,
the global polarization $P_{y}^{G}$ decreases with the collision
energy. However, the circular polarization observables $\langle P_{x}\mathrm{sgn}(Y)\rangle$
and $\langle P_{y}\mathrm{sgn}(Y)\rangle$ are not sensitive to the
collision energy, which is due to that the circular vorticity pattern
has the same magnitude at different collision energies as evidenced
in Fig.~\ref{fig:omega_y}. The local polarization effect along the
longitudinal direction $\langle P_{z}\rangle$ also has a flat energy
behavior. This may be related to that hadron's elliptic flow does
not significantly change with the collision energy \cite{Adamczyk:2015fum}.

Figure \ref{fig:P_phi} shows the results of $\langle P_{x}\mathrm{sgn}(Y)\rangle$,
$\langle P_{y}\mathrm{sgn}(Y)\rangle$, and $\langle P_{z}\rangle$
for the $\Lambda$ hyperons as functions of azimuthal angle $\phi_{p}$
in $20$--$50\%$ central Au+Au collisions at $200$ GeV and Pb+Pb
collisions at $2760$ GeV, where the whole range of $\phi_{p}$ is
divided into 24 bins. We can see that the shapes of $\langle P_{x}\mathrm{sgn}(Y)\rangle$,
$\langle P_{y}\mathrm{sgn}(Y)\rangle$, and $\langle P_{z}\rangle$
are in analogy to $\sin\phi_{p}$, $-\cos\phi_{p}$, and $-\sin(2\phi_{p})$,
respectively, as described by Eq.~(\ref{eq:Fourier}). The features
of three quantities at two collision energies are quite similar. We
have also checked that the harmonic behaviors also exist at energies
$7.7$--$62.4$ GeV. We note that our result for $\langle P_{z}\rangle$
is consistent with the viscous hydrodynamic simulations \cite{Becattini:2017gcx},
while $\langle P_{x}\mathrm{sgn}(Y)\rangle$ and $\langle P_{y}\mathrm{sgn}(Y)\rangle$
are not calculated in that reference. It is worthwhile to point out
that although the global polarization components $P_{x}^{G}$ and
$P_{z}^{G}$ are zero due to the symmetry and $P_{y}^{G}$ is almost
vanishing at $\sqrt{s_{\mathrm{NN}}}=200$ GeV \cite{Adam:2018ivw}
and $2760$ GeV \cite{Timmins:2017gen} due to the reason given in
the above paragraph, the local polarization observables $\langle P_{x}\mathrm{sgn}Y\rangle$,
$\langle P_{y}\mathrm{sgn}Y\rangle$, and $\langle P_{z}\rangle$
are all non-vanishing. We also see in Fig.~\ref{fig:P_energy} that
the magnitude of $\langle P_{y}\mathrm{sgn}Y\rangle$ is larger than
that of $P_{y}^{G}$. Therefore the local polarization effects are
sizable and worthy to be tested in future experiments.

The Fourier coefficients $F_{x}$, $F_{y}$, and $F_{z}$ in Eq.~(\ref{eq:Fourier})
can be extracted from the magnitude of the harmonic behavior in Fig.~\ref{fig:P_phi},
\begin{eqnarray}
F_{x} & = & 2\left\langle P_{x}\mathrm{sgn}(Y)\sin\phi_{p}\right\rangle ,\nonumber \\
F_{y} & = & -2\left\langle P_{y}\mathrm{sgn}(Y)\cos\phi_{p}\right\rangle ,\nonumber \\
F_{z} & = & -2\left\langle P_{z}\sin(2\phi_{p})\right\rangle ,\label{eq:Fourier_coefficient}
\end{eqnarray}
where the averages are taken over $24$ bins of the azimuthal angle.
The results are shown in Fig.~\ref{fig:P_cen} as functions of the
centrality at $\sqrt{s_{\mathrm{NN}}}=200$ GeV for Au+Au and at $2760$
GeV for Pb+Pb collisions. The features of these coefficients are quite
similar at two energies. We see that $F_{x}$ and $F_{y}$ are at
the same magnitude, but there is a difference between them, which
increases with the centrality. This is because the transverse vorticity
loop in non-central collisions should be in an elliptic shape, which
deviates from a prefect circle. We also see that in the most central
collisions $F_{x}$ and $F_{y}$ are non-vanishing, while $F_{z}$
is almost zero. This difference can be understood by the fact that
$F_{z}$ arises from the elliptic flow, which does not exist in central
collisions while $F_{x}$ and $F_{y}$ are generated from the violation
of the longitudinal boost invariance, which exists in both central
and non-central collisions.

\section{\label{sec:Summary}Summary}

We give a systematic analysis on the vorticity structure and the distribution
of $\Lambda$ polarization in heavy-ion collisions. We find that there
are two contributions to the vorticity field: one is from the OAM
along the $-y$ direction giving the global polarization; another
is from the non-uniform expansion of the fireball, which leads to
a circular structure for the transverse vorticity $\bm{\omega}_{\perp}$
and a quadrupole pattern for the longitudinal vorticity $\omega_{z}$
in the transverse plane. The space distribution of the vorticity field
can be probed by the local $\Lambda$ polarization as a function of
the azimuthal angle $\phi_{p}$ and the rapidity $Y$ in momentum
space, which is expected to have harmonic behaviors as in Eq.~(\ref{eq:Fourier}).

For the numerical calculation of the local $\Lambda$ polarization,
we use the string-melting version of the AMPT model. We run the simulations
of Au+Au collisions at energies $\sqrt{s_{\mathrm{NN}}}=7.7$--$200$
GeV and Pb+Pb collisions at $2760$ GeV. We divide all $\Lambda$
hyperons into several bins by their azimuthal angle $\phi_{p}$ and
the sign of $Y$. Then we calculate the average of the polarization
vector for $\Lambda$ in each bin. The results show that $\langle P_{x}\mathrm{sgn}(Y)\rangle$,
$\langle P_{y}\mathrm{sgn}(Y)\rangle$, and $\langle P_{z}\rangle$
have the harmonic behaviors of $\sin\phi_{p}$, $-\cos\phi_{p}$,
and $-\sin(2\phi_{p})$, respectively, which are consistent to the
circular or quadrupole structure of the vorticity field as we expect
from the non-uniform collective expansion of the fireball. These patterns
in the local $\Lambda$ polarization are expected to be tested in
future experiments.
\begin{acknowledgments}
The authors thank Ai-hong Tang and Zhang-bu Xu for insightful discussions.
The authors are supported in part by the Major State Basic Research
Development Program (973 Program) in China under Grants No.~2015CB856902
and No.~2014CB845402 and by the National Natural Science Foundation
of China (NSFC) under Grants No.~11535012 and No.~11720101001.
\end{acknowledgments}

\bibliographystyle{apsrev4-1}
\bibliography{ref}

\end{document}